\documentclass[manuscript,aps]{revtex4}

\begin{document}
\draft

\title{On the formation and the stability of suspended transition metal monatomic chains}

\author{A. Hasmy$^{1,2,*}$, 
L. Rinc\'on$^{1,3}$, 
R. Hern\'andez$^4$, 
V. Mujica$^{1,2}$, M.
M\'arquez$^{1,5}$ and C. Gonz\'alez$^{1}$}

\address{$^1$NIST Center for Theoretical and Computational Nanosciences, National Institute of Standards and Technology,
Gaithersburg, MD 20899, USA}
\address{$^2$INEST Group Postgraduate Program, Philip Morris USA, Richmond, VA 
23234, USA}
\address{$^3$Dpto. de Qu\'\i mica, Facultad de Ciencias, Universidad de 
los Andes, M\'erida-5101, Venezuela}
\address{$^4$Centro de Qu\'\i mica, IVIC, Apdo. 21827, Caracas 1020A, Venezuela}
\address{$^5$Harrington Department of Bioengineering, Arizona State University, Tempe, AZ  85280}

\begin{abstract}
We present a Tight-Binding Molecular Dynamics investigation of the stability, the geometrical and the electronic structure
of suspended monatomic transition metal chains.
We show that linear and stable monatomic chains are formed  at temperature equal or smaller than 500 K for Au, 200 K for Ag and 4 K for Cu.
We also evidence that such stability is associated with the persisting sd orbital hybridization along the chains.
The study highlight fundamental limitations of conductance measurement experiments
to detect these chains in the breaking process of nanowires.
\end{abstract}
\maketitle

Understanding physical properties such as the stability and ubiquity of suspended monatomic chains (SMCs) is an important step to exploit their
abilities
to transport spontaneously spin polarized electrons\cite{Ugarte4,Gillingham} and to support high current densities\cite{CostaKramer}.
In spite of different efforts\cite{Ohnishi,Yanson,Ugarte1,AgraitMD,Bahn,Cadenas3,HSPark,Dreher,Copper2,TAVA},
a theoretical explanation for their relatively high stability and their characteristic linear chain structure remains
elusive\cite{SolerTosatti,SanchezPortal,Fazzio2,prb2007,prl2007}. Moreover, rationalizing the possible formation 
of these chains in 3d and 4d transition metals continues being a challenge
to scientists mainly due to a lack of agreement between
results\cite{Common,Ugarte5,Copper,Contaminants}.

Suspended monatomic chains (SMCs) are typically generated
by electron beam irradiation
of a suspended metal film\cite{Ugarte4,Ugarte1,Cadenas3,Ugarte5,Copper} 
as well through a stretching nanowire procedure that makes use of a scanning tunneling microscope (STM)\cite{Ohnishi,Yanson,AgraitMD}
or a mechanically controllable break junction (MCBJ) technique\cite{Yanson,Common,Contaminants}.
Despite of their popularity, these techniques suffer of inherent limitations that have made the unambiguous interpretation of the data a 
difficult task.
On the one hand,
SMCs formation experiments with electron beam irradiation and STM techniques are based on high-resolution transmission
electron microcopy (HRTEM) observations, which are limited due to
the poor time resolution 
(around thirty pictures per 
second)\cite{Ugarte1} and uncontrollable sample contaminations\cite{legoas}. On the other hand,
MCBJ is based on a not totally justified interpretation of the conductance data generated during the nanowire breaking
process\cite{Common,Contaminants}.
Therefore, computational studies constitute an appealing complement  
that could pave the way for a coherent description of the experimental results.

In this letter, 
the formation and the stability 
of SMCs are investigated
by stretching Au, Ag and Cu nanowires.
Using the tight-binding molecular dynamics technique,
we show that linear and stable SMCs are formed  at temperature equal or smaller than 500 K for Au, 200 K for Ag and 4 K for Cu,
and that such stability is associated with the persisting sd orbital hybridization along the chains.
Additionally, a good agreement is found when numerical histograms of nanowire stretching lengths are
compared to experiments.

For the molecular dynamics simulations, at each iteration step of 2 fs, we calculated the 
electronic structure
by considering the total-energy tight-binding method
introduced by Papaconstantopoulos and Mehl\cite{NRL-TBreview}.
Comparisons with Density Functional Theory calculations revealed that this approximation is able
to describe the SMC formation
process\cite{TAVA,Fazzio2}. 
We used the Verlet algorithm to integrate the equations of motion of each atom.
The temperature $T$ was controlled with a
Langevin thermostat.  The dynamical evolution of the wire under stress was simulated in 
the following way:  for the initial configurations, we considered a metal block containing 68 
atoms obtained from a bulk face centered cubic (fcc) (111) layers oriented along 
the {\it z}-axis and tapered to a narrowing in the middle. The atoms at the top layer were frozen and 
through periodic boundary conditions along the {\it z}-axis, their images were considered to 
form also the bottom layer of the wire. These frozen layers (the real at the top and its 
image at the bottom) determine the bulk support of the wire during the elongation 
process and were used to stretch the structure along the {\it z}-axis by increasing the distance 
between them. The atoms inside these frozen slabs remained frozen during subsequent 
stages of the simulation while the other atoms between them move and rearrange into new
configurations. 
We equilibrated the wire for about 20 ps. Then, as has been done in previous classical
studies\cite{AgraitMD,Bahn,HSPark,Dreher,Copper,HMS}, 
a stretching velocity of 5 m/s was applied. 
The stretching distance is defined as the increased distance 
between the real top layer 
and its image at the bottom respect to the resulting nanowire configuration just before the first atom-contact appears.

Figs. 1a-g show the last stage of a nanowire stretching process when an Au SMC
is formed. 
In all these configurations, 
the bond lengths between the atoms forming the 
chain range from 2.5-2.9 \AA,
in good agreement with experiments\cite{Yanson,Cadenas3}.
Along the chain, we computed the orbital populations based on a Mulliken population analysis, while
the  cohesive energy $E_{coh}$ of an $i$-atom was obtained by calculating
$Tr(\rho H)=\sum{[\rho]_{i\alpha,j\beta}[H]_{j\beta,i\alpha}}$,
where $\rho$ is the density matrix and $H$ the hamiltonian. The above summation is over atom labels $j$ ($\ne i)$ and state labels $\alpha$,$\beta$, and
implicitly includes an integral over {\bf k} as well.
We found that during the nanowire stretching process,
when the monatomic chain gains an extra atom (coming from the 
apexes of the breaking nanowire as seen in Figs. 1a-g),
the electron populations abruptly redistributes within the chain, as reflected by the jumps in the
d-orbital population curves depicted in Fig. 1h. 
A similar effect was found for the s-orbital populations (data not shown), 
in consistency with the sawtoothlike conductance signal observed in experiments during the formation
of Au SMCs\cite{AgraitMD}.
Fig. 1h also shows that depending on the monatomic chain regions,
the d-orbital populations tend to decrease (increase)
in the atoms located at the top and bottom (at the middle) of the chains 
as the stretching distance is increased.
Finally, when this Au monatomic chain (consisting of 6 atoms) breaks, the
the populations abruptly 
increase in the central region
an reach an approximate value of 9.92, a fact that is supported by the drop to zero
of the corresponding $E_{coh}$ (see Fig. 1i). 
Such threshold value corresponds to
the typical d-orbital population for
an atom located in an extreme of an Au monatomic chain. Correlatively, after some relaxation of the broken chain, 
$E_{coh}$  for that atom
jumps from 0 to the corresponding energy value of a single bonded Au atom ($\sim$ -1 eV).

Although the maximum number of atoms contained in a monatomic chain depends on the
stretching realization (as we will discuss below), the above observed abrupt increasing behavior for the d-orbital populations in the middle of the
chain  seems to be independent on the chain length and realization.
The later is evidenced in Fig. 2 for two
Au monatomic chains (black symbols) generated from different stretching process realizations leading to one chain of 4 atoms (black circles) and the other of 5 atoms (black triangles).  
As before, it is observed that $E_{coh}$ decreases when the d-orbital populations become equal to 9.92.
The results in Fig. 2 also show that Ag (blue symbols) and Cu (red symbols) chains of 4 atoms
exhibit a similar behavior.
In these cases, when the chains break, the d-orbital populations result equal to 
9.96 and 9.98 for Ag and Cu, respectively, a fact that may reflect the weaker sd hybridization in these elements.
These results also indicate 
that during the stretching process, 
the loss of the sd orbital hybridization occurs in the lower atomic coordination region, where
the atoms tend to completely fill their d-orbitals.

The stability of SMCs has been also examined here by stopping 
the nanowire stretching process before the chains break. In the molecular dynamics simulations, 
the later is achieved by fixing to zero the stretching velocity applied to the top and the bottom of the nanowire
just when each chain realization reaches
its respective maximum number of atoms. The chains are then allowed to relax at different $T$. 
Fig. 3 shows the d-orbital populations for Au (circles), Ag (squares) and Cu (triangles) at the middle of these chains as a function of time for $T$ around values where the chains become unstable. 
Note that these populations 
fluctuate around a constant value and that the
amplitude of the fluctuations depend on $T$ and the atom type. 
We found that at 4 K, the structure of Au, Ag and Cu monatomic chains remained linear and stable for more than 10 nanosec. 
When $T$ is increased
the amplitude of the fluctuations of the
d-orbital populations increases, leading to an increase of
the probability to break the monatomic chains by reaching the reported d-orbital population threshold values.
The chains break when $T$ is increased to 600 K, 300 K and 100 K in the case of Au, Ag and Cu, respectively.
In addition, for the high $T$ considered in this study, we found that the chains tend to adopt a zigzag structure. 
The later suggests that although this zigzag shape for the chain corresponds to a minimum energy configuration (as predicted by
DFT calculations \cite{SanchezPortal,prl2007}), its experimental observation is not possible because
the required energy to overcome the characteristic potential barrier between the linear and the zigzag chain configuration
is equivalent to a temperature where these chains are unstable. 

In MCBJ experiments, 
the conductance is monitored as a function of the stretching length, showing a
steplike decrease as a nanowire is slowly broken\cite{Yanson,CostaKramer}. 
Since in monovalent transition metals 
each atom located at the nanowire neck 
is expected to
contribute with one quantum unit of conductance
$G_0$ (=2e$^2$/h)\cite{Ohnishi,Ugarte1,monovalent},
the length of the last conductance plateau
(where $G$ is approximately equal to $G_0$)
is usually associated to the chain
length\cite{Yanson,Common,Contaminants}.
A large number of contact breaking cycles
are typically performed in order to build histogram distributions of stretching lengths. At low $T$ and for 5d transition
metals, such histograms exhibit
a series of peaks, 
while a single strong peak is observed in 4d metals. 
The difference between 5d and 4d transition metal histogram distributions of lengths has been used to argue that
only 5d transition metals exhibit SMCs\cite{Common,Contaminants}, 
a fact that contrast with the stability of Ag and Cu
monatomic chains reported in this study.
To investigate the source of such discrepancy, numerical histogram distributions
 of lengths were computed 
by considering different chain formations generated from a hundred numerical realizations of the breaking process
on Au, Ag and Cu nanowires. A very good agreement
between the resulting simulated histograms and those obtained experimentally\cite{Common,Contaminants} is deduced from Fig. 4 (black curves). 
While for T=4 K, Ag and Cu exhibit only a single strong peak,
the Au histogram shows two strong peaks as in the experiments, with a peak-peak separation consistent with the expected bond length of the Au atoms in the chain.
In addition, the
vanishing second peak as $T$ increases observed experimentally for Au\cite{Contaminants} is also predicted by these calculations (see Fig. 4).
This agreement between the simulations and the experiments provides strong confidence on the predictive ability 
of the theoretical approximation considered here.

The simulation data allows a degree of a detailed analysis not possible experimentally,
as is the characterization of the different contributions
of chains containing different number of atoms in the histogram distributions of lengths.
Fig. 4 (colored curves) shows that in the case of Au, Ag and Cu, the first peak
exhibited by these histograms are due to the contributions of
chains with 2 (red curve), 3 (green curve) and 4 (blue curve) atoms, thus indicating that a stretching length (plateau length) of 2 \AA \ can form indistinctively short or long monatomic chains,
depending on each breakage realization
i.e. the particular nanowire evolution process.
Therefore, the stretching length 
required to form a diatomic suspended chain (the two columns of atoms formed just before the
monatomic chain starts to appear) should serve as an indicator of what will happen in the subsequent monatomic chain formation process. 
Thus, in the case of Ag, we found that when the stretching length  associated with the diatomic chain becomes larger than 3 \AA,
a monatomic chain of 4 Ag atoms can be formed by just stretching a length of 2 \AA. However, if the stretching
length corresponding to the diatomic chain formation is smaller than 1 \AA,
a stretching length of more than 4 \AA \ is required to form the same monatomic chain of 4 atoms.
This result leads to the conclusion that
the presence of a single strong maximum around 2 \AA \
in the histogram length distributions cannot be interpreted 
as the absence of monatomic Ag chain formation as previously claimed\cite{Common,Contaminants}, and agrees
with the HRTEM data that provides experimental evidence for the formation of Ag SMCs\cite{Ugarte5}.
The importance of the history of the nanowire breakage process is also observed in the case of Au. 
Figs. 4c and 4d show how the formation of monatomic chains containing 4 Au atoms 
contribute to the high-intensity of the first and second 
maxima of these histogram length distributions. Moreover,
Figs. 1a,g and 4e illustrate how the chain formation process
can differ from one realization to another:
while the structure depicted in Fig. 1a requires
a stretching distance of 5 \AA \  to form an Au chain of 6 atoms, the configuration depicted in the left panel of Fig. 4e
requires only 2.2 \AA  \ to form a similar Au chain of 6 atoms. 

The chain sizes and the critical temperature where these chains lose the stability 
will depend on the specific electronic configuration of each chemical species. The tendency of Au
to produce longer stable monatomic chains can be understood in terms of the strong sd orbital hybridization
that characterize this metal.
The strength of this hybridization can be examined by comparing the orbital populations of an isolated Au atom  
(1 for s and 10 for the d-orbital), 
with the corresponding populations in the bulk phase  (1.5 for s and 9.5 
for d-orbitals), a fact that can be related with the well known strong relativistic effects in the
last row transition metal elements\cite{piko}. 
For other transition metals, these s-orbital (d-orbital) populations are
smaller (higher) as one moves up rows in the periodic table. 

In summary, we show that at low 
$T$ (4 K) Au, Ag and Cu exhibit linear and stable monatomic 
chains, and that these two last elements tend to form shorter chains. Particularly, compared to Au, the tendency of Cu to form shorter chains than 
agrees with a recent result\cite{Copper2}.
Our findings show that the stability
of the chains is associated with the persisting sd orbital hybridization along the chains, which is
affected when the system is heated in an amount that 
depends on the particular metal species. We found that among these three elements, only Au monatomic chains are
highly stable at room temperature.
The present study not only provides a coherent interpretation of the
differences found
in monatomic chain formations between 3d and 4d, with respect to 5d transition
metals\cite{Common,Ugarte5,Copper,Contaminants}, but also is consistent 
with the picture that the sp-character of some atoms like oxygen intercalated in monatomic metal chains
would help the hybridization of d-orbitals, favoring longer chain formations,
as recently suggested in experimental studies on 
controlled oxygen atmosphere\cite{Contaminants}.

We thank T. Allison, J.L. Costa-Kr\"{a}mer, J.J. S\'aenz 
and P.A. Serena for helpful discussions.

*Electronic address: anwar@nist.gov

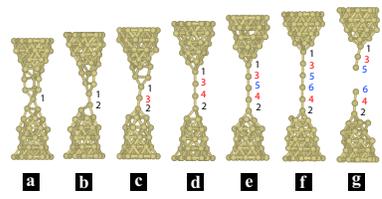
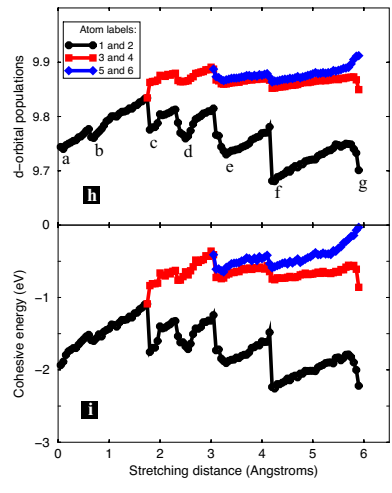
\begin{figure}[ht]
\caption{
(a-g) Schematic
representation of the 
formation of a chain of 6 atoms when a nanowire is stretched. 
(h) d-orbital populations and (i) cohesive
energy as a function of the stretching distance for the different chain regions labeled in figs. (a-g).
} 
\label{fig1}
\end{figure}

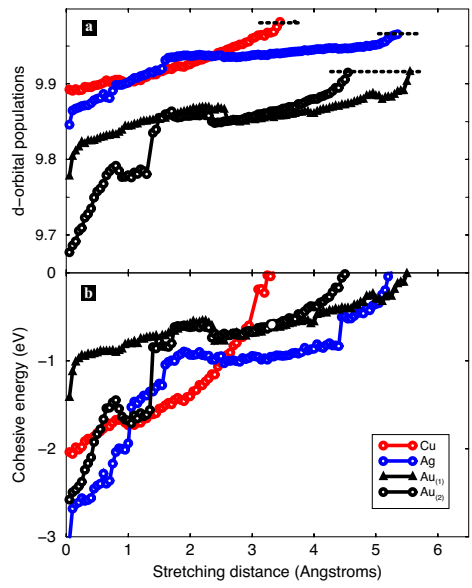
\begin{figure}[ht] 
\caption{
(a) d-orbital populations and (b) cohesive
energy as a function of the stretching distance for atoms located in the central region of Cu, Ag
and Au chains. 
These curves correspond to chains containing a maximum number of 4 atoms (circle symbols) and 5 atoms (triangle symbols).
In (a) the dotted segments denote the threshold value for the d-orbital populations where
the chains break.
} \label{fig2}
\end{figure}

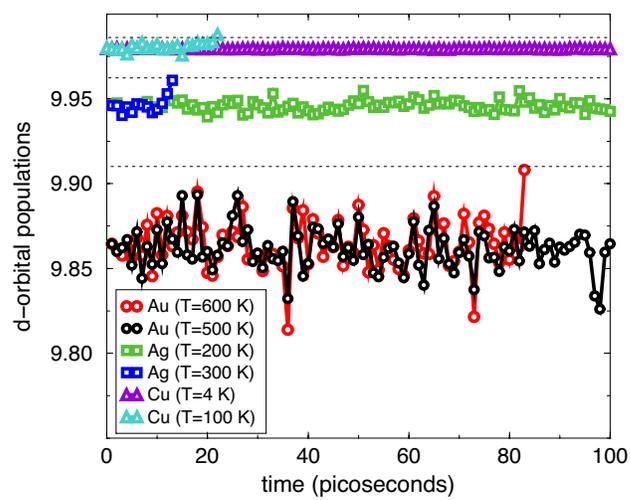
\begin{figure}[ht]
\caption{
Time evolution of the d-orbital populations of the atoms located in the central region of Cu, Ag and
Au chains for different $T$.
Cu and Ag chains contain 4 atoms while the Au chain
contains 6 atoms. The dotted lines depict the typical threshold values where the chain breaks.
} \label{fig3}
\end{figure}

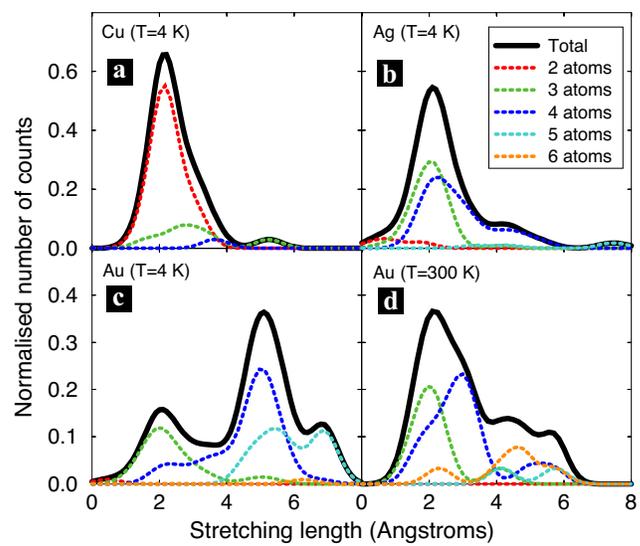
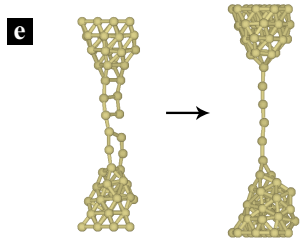
\begin{figure}[ht]
\caption{
(a) Cu, (b) Ag and (c) Au histogram distribution of stretching lengths (black
curves) obtained from a hundred of realizations at $T$ equal to 4 K, and (d) 300 K for Au.
The different contributions in these histograms
of chains containing 2, 3, 4, 5 and 6 atoms are depicted by the colored curves.
(e) Illustrates a situation where a stretching distance of 2.2 \AA \ is enough to form an Au chain of 6 atoms.
} \label{fig4}
\end{figure}

\end{document}